\documentclass[aps, prl,twocolumn,showpacs,preprintnumbers,amsmath,amssymb, a4paper]{revtex4}

\usepackage{graphicx}
\usepackage{dcolumn}
\usepackage{bm}

\begin{document}

\preprint{D. J. Kim \textit{et al.}}

\title{Polarization Relaxation Induced by Depolarization Field in Ultrathin Ferroelectric BaTiO$_3$ Capacitors}

\author{D. J. Kim,$^{1}$ J. Y. Jo,$^{1}$ Y. S. Kim,$^{1}$ Y. J. Chang,$^{1}$ J. S. Lee,$^{1}$
 Jong-Gul Yoon,$^{2}$ T. K. Song,$^{3,}$  }
\email[electronic mail: ]{tksong@changwon.ac.kr}
\author{T. W. Noh$^{1}$}
\affiliation{$^{1}$ReCOE and School of Physics, Seoul National
University, Seoul 151-747, Korea\\
$^{2}$Department of Physics, University of Suwon, Kyunggi-do
445-743, Korea\\
$^{3}$Department of Ceramic Science and Engineering,
Changwon National University, Changwon, Kyungnam 641-773, Korea}

\date{\today}

\begin{abstract}

Time-dependent polarization relaxation behaviors induced by a
depolarization field $E_{d}$ were investigated on high-quality
ultrathin SrRuO$_{3}$/BaTiO$_{3}$/SrRuO$_{3}$ capacitors. The
$E_d$ values were determined experimentally from an applied
external field to stop the net polarization relaxation. These
values agree with those from the electrostatic calculations,
demonstrating that a large $E_{d}$ inside the ultrathin
ferroelectric layer could cause severe polarization relaxation.
For numerous ferroelectric devices of capacitor configuration,
this effect will set a stricter size limit than the critical
thickness issue.

\end{abstract}


\maketitle

With recent breakthroughs in fabricating high-quality oxide films
\cite {Ahn,YSKim1,HNLee}, ultrathin ferroelectric (FE) films have
attracted much attention from the scientific as well as
application points of view. As the FE film thickness $d$
approaches tens of unit cell length, the FE films often show
significantly different physical properties from those of bulk FE
materials. Some extrinsic effects, especially coming from FE film
surfaces and/or interfaces with other materials, could be very
important \cite{Shaw}. For some other cases, intrinsic physical
quantities could play vital roles in determining the unique
properties of ultrathin films.

Many FE-based electronic devices have the capacitor configuration,
where a FE layer is inserted between two conducting electrodes.
Then, polarization bound charges will be induced at the surfaces
of the FE layer, but compensated by free charge carriers in the
conducting electrodes. In real conducting electrodes, however, the
compensating charges will be induced with a finite extent, called
the screening length $\lambda $. This will result in an incomplete
compensation of the polarization charges. Such an incomplete
charge compensation should induce a depolarization field $E_{d}$
inside the FE layer, with a direction opposite to that of the FE
polarization $P$ \cite{Mehta}. Therefore, $E_{d}$ will appear in
every FE capacitor, and its effects will becomes larger with the
decrease of $d$ \cite{Mehta}. (For a FE film without electrodes,
there is no compensation for the polarization bound charge, so the
value of $E_{d}$ will become even larger than that of the FE
capacitor case.) $E_{d}$ has been known to be important in
determining the critical thickness \cite{Junquera} and domain
structure of ultrathin FE films \cite{Kornev,Wu,Fong}, and
reliability problems of numerous FE devices \cite{Kang1,Kang2}.

Recently, using a first principles calculation, Junquera and
Ghosez investigated the critical thickness of BaTiO$_{3}$ (BTO)
layers in SrRuO$_{3}$(SRO)/BTO/SRO capacitor \cite{Junquera}. For
calculations, they assumed that all of the BTO and SRO layers were
fully strained with the SrTiO$_{3}$ substrate. By taking the real
SRO/BTO interfaces into account properly, they showed that $E_{d}$
could make the ferroelectricity vanish for the BTO films thinner
than 6 unit cells, i.e. 2.4 nm \cite{Junquera}. More recently,
using pulsed laser deposition with a reflection high energy
electron diffraction monitoring system, we fabricated high-quality
fully-strained SRO/BTO/SRO capacitors on SrTiO$_{3}$ substrates
with $d$ between 5 and 30 nm \cite{YSKim1,YSKim2}. With a very low
leakage current, we could directly measure their $P$-$E$
hysteresis loops \cite{YSKim1}. In this letter, we report the
time-dependent polarization changes of the ultrathin BTO films. We
find that the net $P$ of the ultrathin BTO films decreases quite
rapidly in time. We will show that the $P$ relaxation should
originate from $E_{d}$. By compensating for $E_{d}$ with an
external potential, we can determine the $E_{d}$ values of the
SRO/BTO/SRO capacitors experimentally. These measured $E_{d}$\
values agree with the values from the electrostatic calculations.
Finally, we will discuss the effect of the $P$ relaxation on a
practical size limitation imposed on actual FE devices.

\begin{figure}[b]
\includegraphics{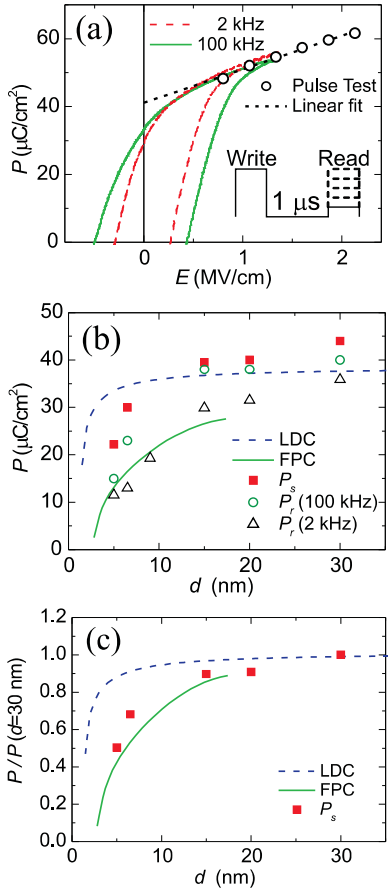}
\caption{(color-online) (a) Upper halves of hysteresis loops for
15 nm thick BTO capacitor at the measurement frequencies 2 and 100
kHz. The values of spontaneous polarization $P_{s}$ were obtained
from linear extrapolation of pulse measurements. (b) The
polarization values obtained from pulse measurements, first
principles calculation (FPC), and Landau-Devonshire calculation
(LDC). (c) Normalized behaviors of LDC, FPC, and $P_{s}$ to the
values of BTO capacitors with 30 nm thickness.}
\label{fig:figure1}
\end{figure}

In our earlier report \cite{YSKim1}, we obtained the
thickness-dependent remnant polarization $P_{r}$ values from the
$P$-$E$ hysteresis loops, measured at 2 kHz in ultrathin FE films
as thin as 5 $\sim $ 30 nm. With further studies on the frequency
dependence of the $P_r$ values in $P$-$E$ hysteresis loops, as
shown in Fig.~1(a) for a 15 nm thick BTO capacitor, we found
differences in the $P_r$ values when the measuring frequency is
varied. These results suggest that the FE domain dynamics should
play an important role for ultrathin FE films, where the FE domain
wall motion is known to be strongly suppressed \cite{Tybell}. Note
that the first principles calculation (FPC) and the
Landau-Devonshire calculation (LDC) do not consider the domain
dynamics, so their predicted polarization values should be called
as spontaneous polarization $P_{s}$.

Since the $P$ value significantly affects the subsequent analysis
of $P$ relaxation, precise determination of $P_s$ values is
necessary. To determine the precise values of $P_{s}$, we applied
pulse trains, which are schematically shown in the inset of
Fig.~1(a) \cite{Smolenskii}. The interval between write and read
pulses was set to 1 $\mu $s to minimize the effects of the $P$
relaxation, and the current responses under the read pulse were
measured. The total amount of charge is obtained by integrating
the current responses in time. The read pulses with different
heights were used to obtain the linear part of the polarization
under an external electric field \ The $P_{s}$ values can be
obtained by extrapolating the linear part of the polarization to
zero electric field. The triangles (black) and circles (green) in
Fig.~1(b) show the $P_{r}$ values measured at 2 and 100 kHz,
respectively. Also, the squares (red) show the $P_{s}$ values from
the pulse test. The solid (green) and dashed (blue) curves show
the theoretical predictions from the FPC \cite{Junquera} and the
LDC \cite{Pertsev}, respectively, which take account of $E_{d}$.
Note that neither of these theories can explain the
thickness-dependence of $P_{s}$ quantitatively. However, it is
known that the FPC predicts systematically somewhat lower bulk
lattice constants compared to real values, so the compressive
stress predicted by the FPC could be smaller than that in the
fully strained sample, resulting in a smaller $P_{s}$. To avoid
this systematic error, we normalized the polarization values to
those of a 30 nm thick BTO capacitor. We found that the
thickness-dependent scaling of $P_{s}$ also follows the FPC
predictions quite well, as shown in Fig.~1(c).

The large difference in $P$ values between the 2 and the 100 kHz
tests indicates that there should be a strong change in the net
$P$ between 10 and 500 $\mu$s. Time-dependent $P$ changes were
investigated by applying two kinds of pulse trains, as shown in
the inset of Fig.~2(a). For the write and the read pulses with the
same (opposite) polarities, the amount of nonswitching (switching)
$P$ can be determined \cite{Kang1}. The difference $\Delta P$,
between the switching and the nonswitching $P$ should be twice as
large as the net $P$. As shown in Fig.~2(a), $\Delta P$ decreases
quite rapidly for the film with $d$ = 15 nm; $\Delta P$ falls to
less than $10\%$ of the $P_s$ value within a relaxation time
$t_{relax}$ of 1000 s. As shown with the solid squares (black) in
Fig.~2(b), $\Delta P$ decay follows a power-law dependence on
$t_{relax}$. Similar power-law decays of $\Delta P$ were observed
for all the BTO films in the thickness range of 5 $\sim $ 30 nm.
Note that such a strong polarization relaxation could pose a
serious problem in capacitor-type ultrathin FE devices.

\begin{figure}[t]
\includegraphics{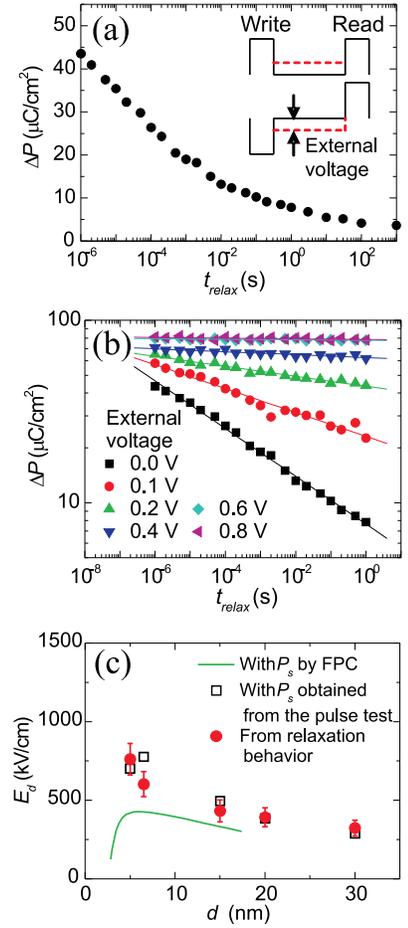}
\caption{(color-online) (a) Polarization relaxation in the 15 nm
thick BTO capacitor. The inset shows the schematics of measuring
relaxation behaviors under an external voltage. (b) Slowing down
of the relaxation behavior under external voltages in the 15 nm
thickness BTO capacitor. (c) Thickness-dependent $E_{d}$ in the
ultrathin BTO capacitors. The $E_{d}$ values obtained from
relaxation behaviors experimentally (red solid circles) and those
from electrostatic calculations with the parameters determined and
measured $P_{s}$ in this work (open squares). The (green) line
indicates the $E_{d}$ from electrostatic calculations with the
polarization values obtained from first principles calculation.}
\label{fig:figure2}
\end{figure}

What is the origin of such strong polarization relaxations? We
thought that they could be closely related to large $E_{d}$
induced inside the BTO films. To verify this idea, we slowed down
the relaxation phenomena by applying an external voltage, as shown
in the inset of Fig.~2(a). The values of the applied external
electric field $E_{ext}$ were obtained by dividing the applied
external voltage by the corresponding film thickness. When the
external field is applied in the opposite direction of $E_{d}$,
the potential gradient inside the FE layer will decrease. Figure
2(b) shows that the slope of the power-law decay becomes smaller,
as $E_{ext}$ increases. Assuming that the depth of the double-well
potential for BTO ferroelectricity can be considered negligible
compared to the effect of $E_{d}$, we approximately determined
experimental $E_{d}$ values from the applied electric field under
which the slope becomes zero. Since $E_{d}$ is proportional to
$P$, the $E_{d}$ value should increase slightly on application of
$E_{ext}$. After correcting this minor contribution, we could
determine the $E_{d}$ values, which are plotted as solid circles
(red) in Fig.~2(c).

From electrostatic calculations on the capacitor geometry, Mehta
\textit{et al.} showed that
\begin{equation}
E_{d}=-\frac{P}{\epsilon _{0}\epsilon _{F}}\left( \frac{2\epsilon _{F}/d}{%
2\epsilon _{F}/d+\epsilon _{e}/\lambda }\right) ,  \label{1}
\end{equation}
where $d$ is the thickness of the FE layer, and $\epsilon _{F}$ and $%
\epsilon _{e}$ are the relative dielectric constants of the FE
layer and the electrode, respectively \cite{Mehta}. To obtain
theoretical $E_{d}$ values for our SRO/BTO/SRO capacitors, we have
to know accurate values of $\epsilon _{e}$, $\lambda $, and
$\epsilon _{F}$. Unfortunately, the reported physical parameter
values in the literature vary \cite{Mehta,Junquera,Black,Dawber}.
Also, we could not find any definite experimental study on
$\epsilon _{e}$.

\begin{figure}[b]
\includegraphics{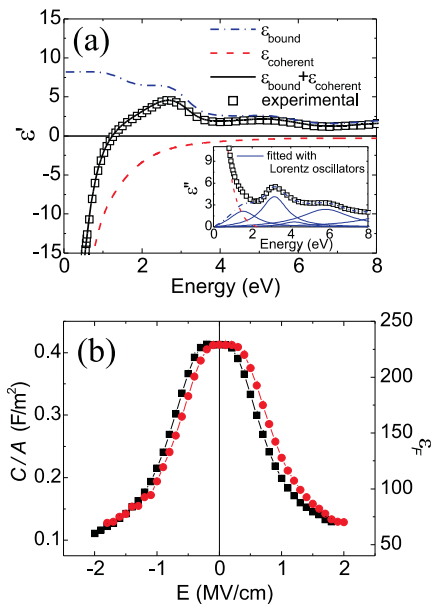}
\caption{(color-online) (a) Frequency-dependent dielectric
functions of SRO. (b) Capacitance-electric field curve of the
SRO/BTO/SRO capacitor with 5 nm thick BTO. } \label{fig:figure3}
\end{figure}

To obtain the value of $\epsilon _{e}$ for an SRO electrode, we
used optical spectroscopy. We measured the optical reflectivity
spectra of epitaxial SrRuO$_{3}$ films (thickness: about 0.5 $\mu
$m) in a wide frequency region between 5 meV and 30 eV and
performed a Kramers-Kronig analysis to obtain the
frequency-dependent dielectric function, $\epsilon (\omega )$
[=$\epsilon ^{\prime }(\omega )+i\epsilon ^{\prime \prime }(\omega
$)]. The details of these measurements and analysis were published
elsewhere \cite{JSLee1, JSLee2}. The open squares in Fig.~3(a) and
the inset show experimental values of $\epsilon ^{\prime }(\omega
)$ and $\epsilon ^{\prime \prime }(\omega )$, respectively. Note
that $\epsilon _{e}$ in Eq.~(1) represents the dielectric response
from the bound charges, namely bound electrons and phonons. Since
SRO is metallic, there should be a large contribution from the
free Drude carriers, which masks the dielectric
response from the bound charges. To obtain $\epsilon _{e}$, we decompose $%
\epsilon (\omega )$ into a free carrier contribution $\epsilon
_{coherent}(\omega )$ and a bound electron contribution $\epsilon
_{bound}(\omega )$ by fitting the experimental $\epsilon (\omega
)$ with a series of Lorentz oscillators, which are displayed as
the dotted (blue) lines in the inset of Fig.~3(a). The dash-dotted
(blue) lines indicate the bound electron contribution. From the
\textit{dc} limit of $\epsilon _{bound}(\omega )$, we could
estimate that the bound electron contribution to $\epsilon _{e}$
is about 8.17. The phonon contribution to $\epsilon _{e}$ was
evaluated in a similar way by analyzing the phonon spectra and
found to be about 0.28 \cite{JSLee1}. Consequently, $\epsilon
_{e}$ is determined to be about 8.45.

Using the carrier density $n_{0}\cong 1.2\times 10^{22}/$cm$^{3}$
of SRO \cite{Shepard}, the experimental value of $\epsilon _{e}$,
and the effective mass of an electron $m_{eff}\cong 7m_{e}$, where
$m_{e}$ is the mass of a free electron \cite{Cao,Okamoto}, we
applied the free electron model and obtained $\lambda =0.8\pm 0.1$
{\AA } \cite{Mehta,Kittel}. We also measured $\epsilon _{F}$ from
the capacitance-electric field $C$-$E$ curves of BTO capacitors.
Figure 3(b) shows the $C$-$E$ curve for the 5 nm BTO capacitor.
The $C$-$E$ curve has the hysteretic behavior typical for a FE
capacitor. The BTO capacitors with 5 $\sim $ 30 nm thickness show
almost the same $\epsilon _{F}$-$E$ curves. The $\epsilon _{F}$
values can vary from 70 to 230 depending on the applied $E$. Since
most of our experiments were performed under a finite applied
field, which corresponds to a value between 1 and 2 MV/cm, the
$\epsilon _{F}$ were estimated to be about 80 \cite{interface}.

With the measured values of $\epsilon _{e}$, $\lambda $, and $\epsilon _{F}$%
, we could estimate the theoretical $E_{d}$ values from Eq.~(1) with the $%
P_{s}$ values obtained from the pulse test. The open squares in
Fig.~2(c) are the theoretical $E_{d}$ values. The solid (green)
line shows the theoretical $E_{d}$ values with the $P_{s}$ values,
obtained from the FPC. These theoretical $E_{d}$ values from the
electrostatic model agree quite well with the experimental $E_{d}$
values, determined from the polarization relaxation. It should be
noted that the $E_{d}$ values are comparable with or even larger
than the measured coercive fields (in our samples, 300 $\sim $ 400
kV/cm). These large $E_{d}$ values can cause $P$ reversal and FE
domain formation, which will result in a reduction of the net $P$
value as time elapses. The fact that two independent
determinations provided nearly the same $E_{d}$ values
demonstrates that \emph{the polarization relaxation behavior
should be dominated by $E_{d}$ inside the FE layer}.

Note that the $E_{d}$-induced $\Delta P$ decay comes intrinsically
from the incomplete compensation of the $P$ charges (due to the
finite screening length of the electrodes) in real conducting
electrode, so that it will inevitably pose a fundamental limit for
most FE device applications using the capacitor configuration.
This limitation should be much more severe than that due to the
critical thickness of the FE ultrathin films \cite{Junquera}. Even
if the FE film is thicker than the critical thickness, it is
feasible that the $E_{d}$-induced $\Delta P$ decay is large enough
to make the net $P$ decrease significantly, resulting in retention
failures for numerous FE devices. As $d$ decreases, $E_{d}$
increases significantly. With the current miniaturization trends
in some FE devices, the large value of $E_{d}$ should play a very
important role in determining the ultimate size limits of FE
devices.

In order to reduce device failure due to the polarization
relaxation, we can try to select better electrode and FE
materials. Noble metals, such as Pt, have been considered better
electrodes because they have high carrier density (resulting in
$\lambda $ values smaller than that of SRO). However, the
$\epsilon _{e}$ values of typical noble metals are much smaller
than that of SRO, i.e. 8.45 \cite{Ehrenreich}, so $E_{d}$ in
capacitors with noble metal electrodes can be large. For example,
$E_{d}$ in the range of 500 $\sim $ 900 kV/cm is expected for a 15
nm thick BTO film with noble metal electrodes (typically, $\lambda
=0.4\sim 0.5$ {\AA }, $\epsilon _{e}=2\sim 4$). Thus, the
$E_{d}$-induced $P$ relaxation for the ultrathin BTO capacitors
with the noble metal electrodes could be at least equal to or
worse than that with SRO electrodes. Proper FE material selection
can be another option. Since PbTiO$_{3}$ is known to have a much
deeper double-well potential than that of BTO
\cite{Pertsev,Cohen}, the $P$ relaxation should occur at a much
lower rate even with the same value of $E_{d}$. Optimization of FE
materials should be of great importance for the improvement of
ultrathin film nanoscale FE device performances.

In summary, we demonstrated that the depolarization field inside
the ferroelectric film could cause a severe polarization
relaxation. By slowing down the relaxation under an external
field, we could determine the depolarization field in a real
capacitor of ultrathin SrRuO$_{3}$/BaTiO$_{3}$/SrRuO$_{3}$
experimentally, which result is in good agreement with
electrostatic calculations. Our investigation demonstrates that
the depolarization field originates from intrinsic properties of
electrode material such as the finite screening length and that
the depolarization field should play an important role in domain
dynamics in ultrathin FE films. The polarization relaxation due to
the depolarization field could pose a serious size limitation for
ultrathin ferroelectric devices.

The authors thank Prof. Sug-Bong Choe in Seoul National University
for valuable discussions. This work was financially supported by
the Korean Ministry of Science and Technology through the Creative
Research Initiative program and by KOSEF through CSCMR.

\newpage


\begin{thebibliography}{10}


\bibitem{Ahn}  C. H. Ahn, K. M. Rabe, and J.-M. Triscone, Science \textbf{303}, 488 (2004).

\bibitem{YSKim1}  Y. S. Kim, D. H. Kim, J. D. Kim, Y. J. Chang, T. W. Noh,
J. H. Kong, K. Char, Y. D. Park, S. D. Bu, J.-G. Yoon, and J.-S.
Chung, Appl. Phys. Lett. \textbf{86}, 102907 (2005).

\bibitem{HNLee}  H. N. Lee, H. M. Christen, M. F. Chisholm, C. M. Rouleau,
and D. H. Lowndes, Nature \textbf{433}, 395 (2005).

\bibitem{Shaw}  T. W. Shaw, S. Trolier-McKinstry, and P. C.
McIntyre, Annu. Rev. Mater. Sci. \textbf{30}, 263 (2000).

\bibitem{Mehta}  B. B. Mehta, B. D. Silverman, and J. T. Jacobs, J. Appl.
Phys. \textbf{44}, 3379 (1973).

\bibitem{Junquera}  J. Junquera and P. Ghosez, Nature \textbf{442}, 506
(2003).

\bibitem{Kornev}  I. Kornev, H. Fu, and L. Bellaiche, Phys. Rev. Lett.
\textbf{93}, 196104 (2004).

\bibitem{Wu}  Z. Wu, N. Huang, Z. Liu, J. Wu, W. Duan, B.-L. Gu, and X.-W.
Zhang, Phys. Rev. B \textbf{70}, 104108 (2004).

\bibitem{Fong}  D. D. Fong, G. B. Stephenson, S. K. Streiffer, J. A.
Eastman, O. Auciello, P. H. Fuoss, and C. Thompson, Science
\textbf{304}, 1650 (2004).

\bibitem{Kang1}  B. S. Kang, J.-G. Yoon, T. W. Noh, T. K. Song, S. Seo, Y.
K. Lee, and J. K. Lee, Appl. Phys. Lett. \textbf{82}, 248 (2003).

\bibitem{Kang2}  B. S. Kang, J.-G. Yoon, D. J. Kim, T. W. Noh, T. K. Song,
Y. K. Lee, J. K. Lee, and Y. S. Park, Appl. Phys. Lett.
\textbf{82}, 2124 (2003).

\bibitem{YSKim2}  Y. S. Kim, J. Y. Jo, D. J. Kim, Y. J. Chang, J. H. Lee, T.
W. Noh, T. K. Song, J.-G. Yoon, J.-S. Chung, S. I. Baik, Y.-W.
Kim, and C. U. Jung, cond-mat/0506495 (2005).

\bibitem{Tybell}  T. Tybell, P. Paruch, T. Giamarchi, and J.-M. Triscone,
Phys. Rev. Lett. \textbf{89}, 097601 (2002).

\bibitem{Smolenskii}  G. A. Smolenskii, V. A. Bokov, V. A. Isupov, N. N.
Krainik, R. E. Pasynkov, and A. I. Sokolov, \textit{Ferroelectrics
and Related Materials} (Gordon and Breach, New York, 1984) pp4-5.

\bibitem{Pertsev}  N. A. Pertsev, A. G. Zembilgotov, and A. K. Tagantsev,
Phys. Rev. Lett. \textbf{80}, 1988 (1998).

\bibitem{Black}  C. T. Black and J. J. Welser, IEEE Trans. Electron Devices
\textbf{46}, 776 (1999).

\bibitem{Dawber}  M. Dawber, P. Chandra, P. B. Littlewood, and J. F. Scott,
J. Phys.: Condens. Matter \textbf{15}, L393 (2003).

\bibitem{JSLee1}  J. S. Lee, Y. S. Lee, T. W. Noh, S. Nakatsuji, H.
Fukazawa, R. S. Perry, Y. Maeno, Y. Yoshida, S. I. Ikeda, J. Yu,
and C. B. Eom, Phys. Rev. B \textbf{70}, 085103 (2004).

\bibitem{JSLee2}  J. S. Lee, Y. S. Lee, T. W. Noh, K.Char, J. Park,
S.-J. Oh, J.-H. Park, C. B. Eom, T. Takeda, and R. Kanno, Phys.
Rev. B \textbf{64}, 245107 (2001).

\bibitem{Shepard}  M. Shepard, S. McCall, G. Cao, and J. E. Crow, J. Appl.
Phys. \textbf{81}, 4978 (1997).

\bibitem{Cao}  G. Cao, S. McCall, M. Shepard, J. E. Crow, and R. P. Guertin,
Phys. Rev. B \textbf{56}, 321 (1997).

\bibitem{Okamoto}  J. Okamoto, T. Mizokawa, A. Fujimori, I. Hase, M. Nohara,
H. Takagi, Y. Takeda, and M. Takano, Phys. Rev. B \textbf{60},
2281 (1999).

\bibitem{Kittel}  C. Kittel, \textit{Introduction to Solid State Physics},
6th ed. (John Wiley \& Sons, Inc., New York, 1996) pp280-282.

\bibitem{interface}
Note that the charge compensation in the finite screening length
in electrodes might result in space charge capacitance in series
with ferroelectric capacitance. For the case of the 5 nm thick
film, $\epsilon _{F}$ should be increased to 112 from 80 with the
correction. However, this increase of $\epsilon _{F}$ does not
change $E_{d}$ so much; from 708 to 650 kV/cm (about 8\%). Because
this is the most serious case, corrections are not necessary for
thicker films.

\bibitem{Ehrenreich}  H. Ehrenreich and H. R. Phillipp, Phys. Rev. \textbf{%
128}, 1622 (1962).

\bibitem{Cohen}  R. E. Cohen, Nature \textbf{358}, 136 (1992).


\end{thebibliography}
\end{document}